# Automatic Generation of RAMS Analyses from Model-based Functional Descriptions using UML State Machines

Christof Kaukewitsch, Henrik Papist, Marc Zeller, Martin Rothfelder

Siemens AG, Corporate Technology, Munich

*SUMMARY & CONCLUSIONS*

In today's industrial practice, safety, reliability or availability artifacts such as fault trees, Markov models or FMEAs are mainly created manually by experts, often distinctively decoupled from systems engineering activities. Significant efforts, costs and timely requirements are involved to conduct the required analyses. In this paper, we describe a novel integrated model-based approach of systems engineering and dependability analyses. The behavior of system components is specified by UML state machines determining intended/correct and undesired/faulty behavior. Based on this information, our approach automatically generates different dependability analyses in the form of fault trees. Hence, alternative system layouts can easily be evaluated. The same applies for simple variations of the logical input-output relations of logical units such as controllers. We illustrate the feasibility of our approach with the help of simple examples using a prototypical implementation of the presented concepts.

## 1  INTRODUCTION

Safety, reliability or availability artifacts are essential with respect to the homologation of complex systems or solutions as well as with respect to the fulfillment of corresponding contractual obligations in general.

Nowadays, such dependability artifacts for technical systems or solutions are often distinctively decoupled from systems engineering activities. Moreover, corresponding artifacts are usually generated manually by reliability, availability, maintainability or safety (RAMS) experts and related teams subsequently to design decisions. Significant efforts, costs and time for experts are required to generate the needed dependability artifacts especially in case of complex system designs. Moreover, the increasing complexity of modern system architectures such as for instance in Cyber-Physical Systems (CPS) that reconfigure during run-time and the predominant importance of short time-to-market entail the need for model-based and automated methods in systems engineering as well as in the dependability realm.

The method presented in this paper describes a new integrated approach of systems engineering and dependability analyses. It involves state-of-the-art component-based system design using standard Unified Modeling Language (UML)[1] or System Modelling Language (SysML)[2] methodology. The behavior of system components is specified by UML state machines determining intended/correct and undesired/faulty behavior. Given this information, the user interaction is limited to defining the failure criterion, the so-called top event of a fault tree, by choosing proposed state combinations at the system boundary. This allows automatic generation of fault trees for e.g. reliability and safety similarly by simply adapting the top event definition. In case of system modifications due to changed topology or functionality automated fault tree generation is continued given the underlying failure definitions persist. Consequently, a huge advantage of this method is that the fulfillment of dependability-related requirements can be continuously verified and assured. This even holds in case of CPS that reconfigure during run-time. In addition, efforts once spent for the definition of components can easily be reused in different contexts and by different users.

The rest of this paper is structured as follows: First, we present related work in section 2. Then, we present in section 3 our approach to automatically generate RAMS analyses based on the functional model in form of UML state charts. Section 4 provides a case study to demonstrate the feasibility of our approach. Section 5 summarizes this paper.

## 2  RELATED WORK

Nowadays, in industrial practice dependability analyses for technical systems or solutions (which includes human activities for system operation and maintenance) such as Failure Modes and Effects Analysis (FMEA and their derivatives), Fault Trees or Markov Chains typically are generated manually by experts or related teams. Significant efforts, costs and timely requirements are involved especially in case of complicated or challenging applications. This also includes efforts to synchronize between system designers and RAMS experts and comprises the risk of misunderstanding and false interpretation resulting in inappropriate dependability analyses.

In previous work, efforts have been made in order to partially automate the generation of safety analyses. [1,2,3,4,5]

---

[1] https://www.omg.org/spec/UML/

[2] https://sysml.org/

present approaches to generate FMEA tables from system models. In [6,7,8,9,10] fault tree models are generated from system models to perform safety analyses. To construct the analyses the system models are often annotated with failure propagation models [11,12,13,14,15]. However, all these approaches focus on the system architecture as input and often require manual modeling of the failure propagation. Preparation efforts are very high in order to be able to generate RAMS analyses.

Other previous approaches deal with the automated generation of fault trees from behavioral specification in form of state machines [16,17,18,19] or mode automat [20,21]. Approaches like Altarica [22] are also available as software tools for model-based safety analysis (e.g. Simfia Neo). However, in all of these approaches the generation of fault trees requires manual annotation of failure modes by RAMS experts. Moreover, all previous approaches focus on the generation of safety analyses whereas our approach presented in this paper not only allows the generation of safety analyses but also of reliability or availability analyses. Hence, manual effort to create the model is only required once, and several RAMS analyses can be generated based on this model. Moreover, component models can be stored in a library and be reused for the modeling of new systems.

### 3 AUTOMATIC GENERATION OF RAMS ANALYSES

Model-based system engineering got a lot of attention in the last decades and paved its way into industrial practice. System engineers describe the technical specification in a model-based way with a standardized generic functional description (e.g. with UML or SysML). This description includes the components of the system and its relations using input and output port definitions, which are interconnected, e.g. by using SysML Internal Block Diagrams (see Fig. 1). The functional behavior of the system in the approach presented in this paper is described by modeling each component using UML state machines. The states of a state machine represent the nominal behavior of the component but also the failure modes of the component by dedicated states (so-called failure states). The behavior of the component in each state is defined using the Object Constraint Language (OCL), see Fig. 2.

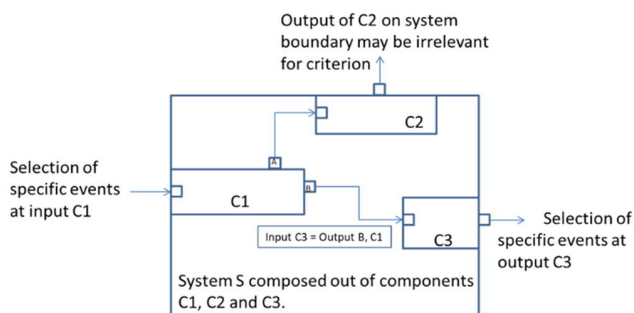

*Figure 1 – Exemplary System S composed of 3 components*

For every state of the state machine the relevant output state, which may depend on input states, is defined. Output states of a component at the system boundary (e.g. component C3 in Fig. 1) depend on the values or states at the inputs of this component and its UML state machine. The inputs to the component depend on the UML state machines of the connected components and their input values and so forth. Some UML failure states of the component may be modeled as not depending on input values. Hence, the output values only depend on the failure mode itself. This may be adequate for certain failure modes e.g. with inherent failure monitoring that reports a faulty state on its output in case an internal failure such as for instance a memory defect has been detected.

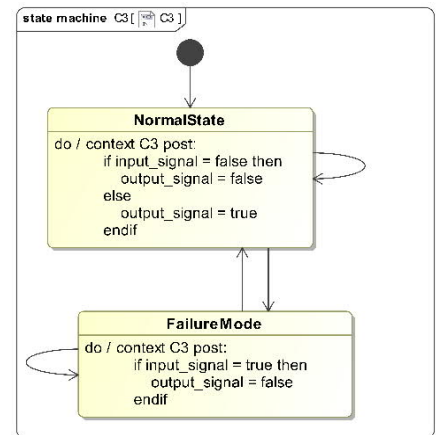

*Figure 2 – Exemplary UML state machine for component C3*

For instance, one state of the state machine of a cable segment may relate to a short circuit failure mode on the output side. This consequently leads to a very low resistance on the output of this component independently from the signals on the input side (reasonable technical input states supposed). In case of logic devices or units a useful generic approach is to predefine general behavior and known failure modes by means of corresponding states. Then it is up to the user to add specific logical input-output dependencies according to a pre-specified rule set to implement the intended functionality.

It is supposed that the generic description of the component or sub-system reflects certain standardized implementation concepts or architectural patterns, e.g. with respect to sensor circuitry, signaling or communication protocols which facilitates system development. The system designer can select, detail or enrich the relevant component functionality with the help of these architectural patterns in the course of the system definition. Therefore, the system engineer will only encounter a one-time effort in order to prepare the system or solution for the generation of the different, relevant RAMS analyses. Moreover, the defined failure behavior can be delivered by the supplier together with the component as digital twin. This data can be reused each time the system component is implemented in a solution.

Based on the information about the system structure and the functional behavior of each component, it is possible to generate RAMS analyses using our approach comprises the following steps:
1) The user selects the relevant states (input and output states) at the system boundary to define the failure criterion. In

case of a "false negative failure" this could for instance be an output state or several output states indicating the absence of a to-be-detected hazard while in fact the corresponding input to the system signals the hazard being present. The set of possible output states can be easily derived in case the corresponding component only entails states with explicit output values. If output states are present, which depend on their input states, it is required to involve upstream components in the analysis. Hence, even a full system analysis may be required.

2) Once the failure criterion is set the algorithm to generate the fault tree can be started: Failure modes that cannot occur according to the selection of input states at the system boundary have to be neglected for further analyses. If there are UML failure states of the component at the system boundary that are independent from input states (technical reasonable values supposed) and that result in one of the selected system output states the corresponding failure modes are integrated into the fault tree under an OR-gate. UML states not depending on input values and not resulting in an output value that represents one of the system output states according to the chosen failure criterion can be ignored for further analyses.

3) UML states depending on input states must be further evaluated in order to determine whether the resulting states can match the defined output states according to the failure criterion. Therefore, it is required to include the UML state machines of the upstream components. If corresponding failure modes directly lead to the selected output states, these also have to be included in the fault tree under the OR gate. Again, evaluation is straight-forward for failure states with explicit output values that do not depend on input values.

4) If the implemented logic of a component requires a certain combination of inputs to result in the selected output states corresponding failure states have to be integrated into the fault tree under an AND gate.

5) The algorithm is terminated once the system boundary is reached and the fault tree is displayed.

Since in this algorithm failure modes (i.e. states which must be evaluated) can be neglected according to the selection of the criteria, it scales good even in case of an increasing number of components and system states.

## 4 CASE STUDY

In this case study, we demonstrate the feasibility of our approach by means of a fire detection system model comprising three equal fire detectors (FD1, FD2, FD3), three equal cable segments (C1, C2, C3) and one controller unit evaluating the fire detectors' signals ("Alarm", if a fire is detected via infrared signal, and otherwise "no Alarm") transmitted by cable segments (see Fig. 3). The so-called "safe logic" implemented in CPU1 triggers an alarm unless the three inputs to CPU1 signal "no Alarm". Alternatively, the controller unit can be implemented in CPU2 using a so-called "voting logic" which triggers the alarm, if the majority of inputs (two in this example) to CPU2 signal "Alarm".

For each of the components in this exemplary system, a UML state machine is defined. The fire detector is modelled with four states. The initial state is the "*NormalState*" in which the component is correctly working. The behavior of the detector is defined in this state using OCL. Moreover, the fire detector has 3 different failure modes (FM): *MissedAlarm*, *FalseAlarm*, and *Internal-Failure-Detected* (see Fig. 4). The latter one is independent of inputs states while the other failure modes can only contribute to the component behavior in case an infrared signal is present (*MissedAlarm*) or absent (*FalseAlarm*) at the system inputs.

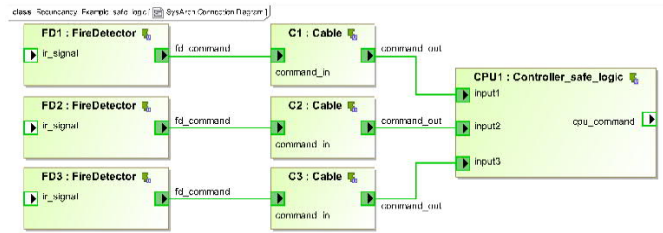

*Figure 3 – Fire Detection System composed of different components*

The cable component has besides the "*NormalState*" two failure modes: "*Open-Circuit*" and "*Short-Circuit*" (see Fig. 5).

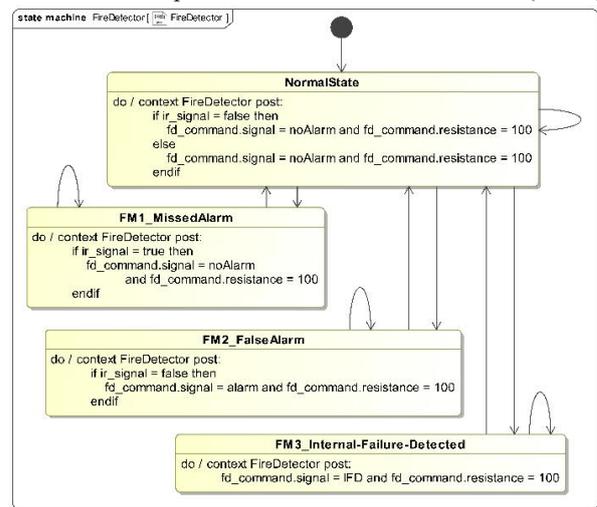

*Figure 4 – UML state machine of the Fire Detector component*

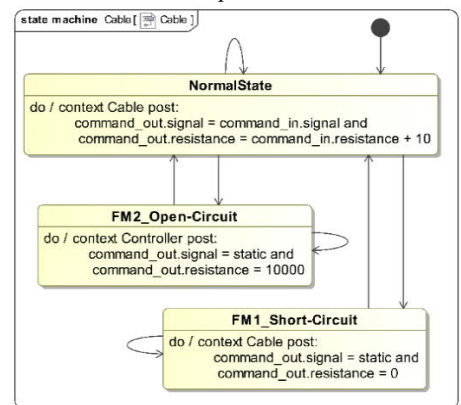

*Figure 5 – UML state machine of the Cable component*

The controller component consists of the "*NormalState*" as well as the failure states "*Internal-Failure-Detected*", "*No_Output*", and "*Unknown_Signal*". These failure states are the same in CPU1 and CPU2. The two variants only differ w.r.t. the logic defined in the NormalState (see Fig. 6 and Fig. 7).

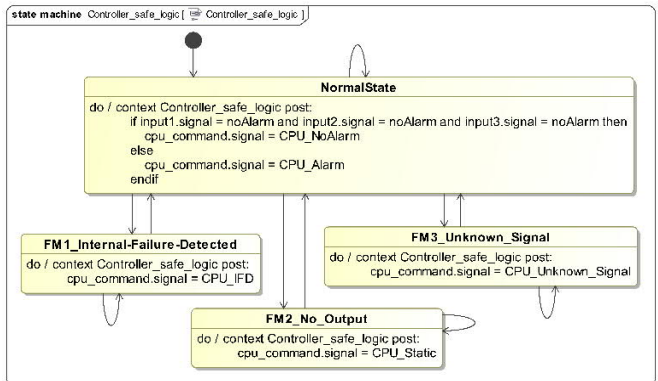

*Figure 6 – UML state machine of controller CPU1 with safe logic*

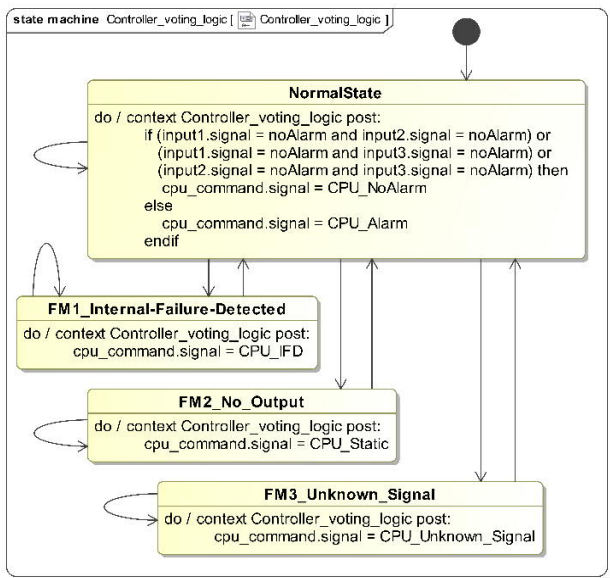

*Figure 7 – UML state machine of controller CPU2 with voting logic*

Based on the description of the system structure in form of the SysML Internal Block Diagram IBD (in Fig. 3) and the definition of the behavior of each component in form of UML state machines (see Fig. 4 to Fig. 7), it is possible to automatically generate RAMS analyses for the exemplary system for different failure criteria as described in Section 3.

### 4.1 Variant 1: False Negative (Safety Analysis)

In this variant, CPU1 is used as the controller in the fire detector systems. Moreover, all input states to the fire detectors are set to infrared signal being present indicating a potential fire, while the output state selected for CPU1 is "*CPU_NoAlarm*". This failure criterion relates to a potential safety-critical event - a missed alarm in case of a fire.

The resulting fault tree for this failure criterion is displayed in Fig. 8. As expected for the safe-logic implementation, a safety-critical failure "*CPU_NoAlarm*" only occurs, if all three fire detectors (FD1, FD2, FD3) simultaneously miss the infrared signal (failure state "*MissedAlarm*").

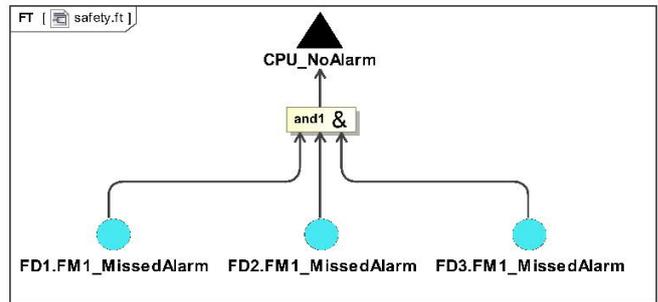

*Figure 8 – Generated fault tree for the safety analysis of the example with CPU1 (safe logic)*

### 4.2 Variant 1: False Positive (Availability Analysis)

An availability analysis can also be derived by an adaption of the failure criterion. In this case, the output state selected for CPU1 is "*CPU_Alarm*". This failure criterion relates to all failures leading to a false alarm, i.e. an alarm occurs without a fire and hence without infrared radiation being present which constitutes an undesired source of unavailability. Therefore, all input states to the fire detectors are set to infrared signal not being present.

The resulting fault tree for the availability analysis is depicted in Fig. 9. It contains no failure mode of CPU1, since normal behavior is required to output the "*CPU_Alarm*" signal. Also, it does not comprise the "*missed alarm*" failure modes of the fire detectors, since the input state definition excludes infrared radiation. All other failure modes ("*FalseAlarm*", "*Internal-Failure-Detected*") are included to the fault tree for availability analysis, since any deviation from the "*no_Alarm*" signal triggers "*CPU_Alarm*" according to the safe logic implemented.

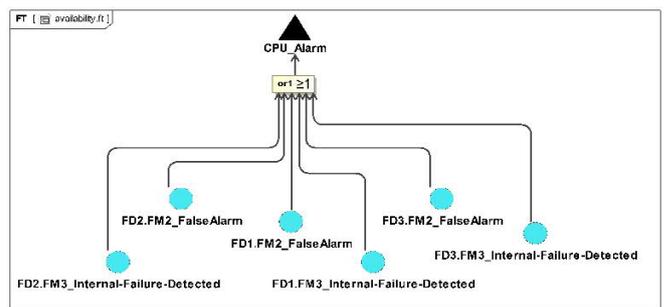

*Figure 9 – Generated fault tree for availability analysis of the example with CPU1 (safe logic)*

### 4.3 Variant 2: False Negative (Safety Analysis)

In this variant, CPU2 is used as the controller in the fire detector systems. Again, all input states to the fire detectors are set to infrared signal being present indicating a potential fire, while the output state selected for CPU1 is "CPU_NoAlarm".

*Figure 10 – Generated fault tree for the safety analysis of*

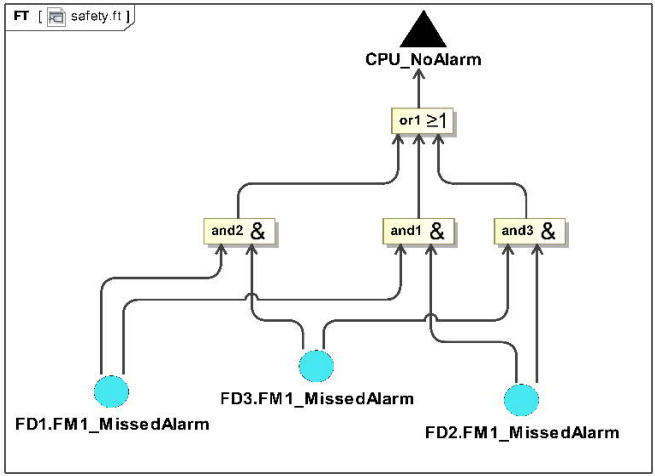

*the example with CPU2 (voting logic)*

The resulting fault tree for this failure criterion is depicted in Fig. 10. In contrast to the safe-logic implementation of CPU1 (cf. Fig. 8), a safety-critical failure "*CPU_NoAlarm*" occurs, if two out of three fire detectors (FD1, FD2, FD3) simultaneously miss the infrared signal. This behavior is respected and represented in the automatically generated fault tree by the OR gate and the three AND gates (2 out of 3 logic).

*4.4 Variant 2: False Positive (Availability Analysis)*

Again, an availability analysis is generated for the variant with CPU2 as the controller using the same criteria as described in Section 4.2. The automatically generated fault tree is depicted in Fig. 11. It also includes a voting logic, since due to the voting logic implemented in CPU2 "*CPU_Alarm*" is triggered, if two fire detectors fail to send the "*no_Alarm*" signal (in any combination of failures).

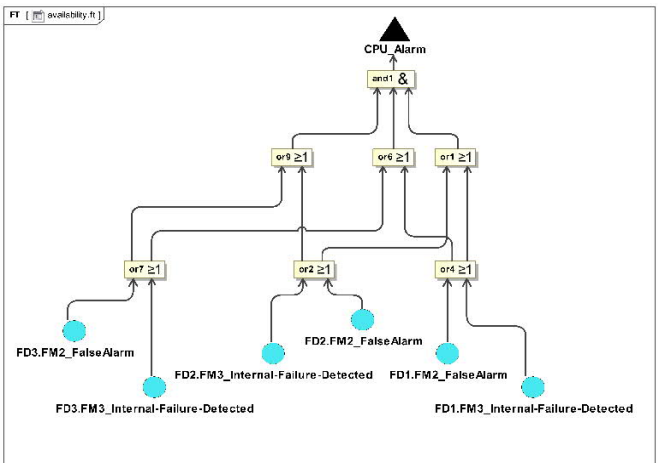

*Figure 11 – Generated fault tree for the availability analysis of the example with CPU2 (voting logic)*

## 5 CONCLUSIONS & FUTURE WORK

In this paper, we present a novel integrated approach of systems engineering and dependability analyses. Based on the functional description of system components our method generates dependability analyses such as fault tree analyses automatically. Once the system model is implemented the user's interaction is limited to defining the failure criteria for the top events of the required fault trees by choosing corresponding state combinations from proposed states at the system boundary. The fault trees are then automatically generated and displayed and can be evaluated with corresponding analysis software. Alternative system compositions as well as different logical functionality for logical units can easily be evaluated using this approach, while the expert has to define the failure behavior only once. Currently our approach is limited to systems with simple logical dependencies between inputs and outputs. Implementation examples support the feasibility of the approach. Future work will be the enhancement of our approach to represent time-dependent behavior and to generate FMEA sheets automatically for RAMS analyses. Also, the tooling capabilities will be extended correspondingly. We plan to evaluate our approach in detail using more complex system configurations.

*ACKNOWLEDGMENT*

Parts of the work presented in this paper were created in the context of the DEIS Project (Dependability Engineering Innovation for CPS), which is funded by the European Commission (Grant Agreement No. 732242).

*BIOGRAPHIES*

Christof Kaukewitsch
Siemens AG, Corporate Technology
Otto-Hahn-Ring 6, Munich, 81739, Germany

e-mail: christof.kaukewitsch@siemens.com

Christof Kaukewitsch works as a RAMS engineer at Siemens AG, Corporate Technology, in Munich since 2009 and has worked on topics related to system integration and service and maintenance before. His research interests refer to model-based reliability and safety analyses of complex systems. He graduated as Electrical Engineer from the Technical University of Darmstadt in 1993 and received a Master of Science degree in Systems Engineering at the Clausthal University of Technology in 2013.

Henrik Papist
Siemens AG, Corporate Technology
Otto-Hahn-Ring 6, Munich, 81739, Germany

e-mail: henrik.papist@siemens.com

Henrik Papist studied mechanical engineering at the Technical University Hamburg (TUHH) and graduated in 2017 as B.Sc. Since 2017 he is studying for a master's degree in aerospace technology at the Technical University Munich (TUM).

Dr. Marc Zeller
Siemens AG, Corporate Technology
Otto-Hahn-Ring 6, Munich, 81739, Germany

e-mail: marc.zeller@siemens.com

Marc Zeller works as a research scientist at Siemens AG, Corporate Technology, in Munich since 2014. His research interests are focused on the model-based safety and reliability engineering of complex software-intensive embedded systems. He studied Computer Science at the Karlsruhe Institute of Technology (KIT) and graduated in 2007. He obtained a PhD from the University of Augsburg in 2013 for his work on self-adaptation in networked embedded systems at the Fraunhofer Institute for Embedded Systems and Communication Technologies ESK in Munich.

Martin Rothfelder
Siemens AG, Corporate Technology
Otto-Hahn-Ring 6, Munich, 81739, Germany

e-mail: martin.rothfelder@siemens.com

Martin Rothfelder received his diploma in Electrical Engineering from Ruhr-University Bochum in 1991. He started as functional safety assessor for TÜV Rheinland. 1996 he joined Siemens. Now he heads the Research Group Dependability Analysis & Management. Martin Rothfelder has long-term industrial experience in safety management (rail, automotive, industrial controls) and is author of many publications in this area. His current research focusses on Model-based Reliability & Safety Engineering, and V&V of Autonomous Systems.